\documentclass[12pt]{article}
\usepackage[a4paper,margin=2cm]{geometry}
\usepackage{latexsym}
\usepackage{graphicx}
\usepackage{mathptmx}
\usepackage{authblk}
\usepackage{amsmath}
\usepackage{bm}
\usepackage{amsfonts}
\usepackage{amssymb}
\usepackage{amsbsy}
\usepackage{amsthm}
\usepackage[mathscr]{euscript}
\usepackage{amsmath}
\usepackage{graphicx}
\usepackage{url} 
\usepackage{changepage}
\usepackage[utf8x]{inputenc}
\usepackage{textcomp,marvosym}
\usepackage{fixltx2e}
\usepackage{amsmath,amssymb}
\usepackage{nameref,hyperref}
\usepackage{authblk}
\usepackage{algorithm}
\usepackage{algpseudocode}
\usepackage{dsfont}
\usepackage{xcolor}
\usepackage{enumitem}
\usepackage{caption,setspace}
\usepackage{pdflscape}
\usepackage{graphicx}
\usepackage{placeins}
\usepackage{mathtools}
\usepackage{appendix}
\usepackage{extarrows}
\usepackage{natbib}
\usepackage{amsmath, amssymb, graphicx, psfrag}
\usepackage{color}
\usepackage{url}

\usepackage{physics}
\usepackage{siunitx}
\usepackage{bm}

\usepackage{hyperref}

\newtheoremstyle{wsc}
{3pt}
{3pt}
{}
{}
{\bf}
{}
{.5em}
{}

\theoremstyle{wsc}

\newcommand{\CondEsub}[3]{\mathbb{E}_{#3}\left[#1 \middle| #2\right]}
\newcommand{\CondVarsub}[3]{\text{Var}_{#3}\left[#1 \middle| #2\right]}
\newcommand{\bvec}[1]{\mathbf{#1}}
\DeclareMathOperator*{\argmax}{argmax}
\DeclareMathOperator*{\argmin}{argmin}
\DeclareMathOperator*{\diag}{diag}

\newcommand{\pddydxdz}[3]{\frac{\partial^2 #1}{\partial {#2}\partial {#3}}}
\newcommand{\Esub}[2]{\mathbb{E}_{#2}\left[#1\right]}

\newcommand{\bb}{\mathbf{b}}

\newcommand{\bzero}{\mathbf{0}}

\newcommand{\bbeta}{\boldsymbol{\beta}}

\newcommand{\bepsilon}{\boldsymbol{\epsilon}}

\begin{document}

%
%

\title{Robust simulation design for generalised linear models in conditions of heteroscedasticity or correlation}

\author[1]{Andrew~Gill\footnote{To whom correspondence should be addressed. E-mail: andrew.gill@defence.gov.au}}
\author[2,3]{David~J. Warne}
\author[4]{Antony~M. Overstall}
\author[2]{Clare McGrory}
\author[2,3]{James~M. McGree}

\affil[1]{Joint and Operations Analysis Division, Defence Science and Technology Group, Edinburgh, Australia}
\affil[2]{School of Mathematical Sciences, Queensland University of Technology, Brisbane, Australia}
\affil[3]{Centre for Data Science, Queensland University of Technology, Brisbane, Australia}
\affil[4]{Mathematical Sciences, University of Southampton, Southampton, UK}
\date{\today}
\maketitle
\begin{abstract}
A meta-model of the input-output data of a computationally expensive simulation is often employed for prediction, optimization, or sensitivity analysis purposes. Fitting is enabled by a designed experiment, and for computationally expensive simulations, the design efficiency is of importance. Heteroscedasticity in simulation output is common, and it is potentially beneficial to induce dependence through the reuse of pseudo-random number streams to reduce the variance of the meta-model parameter estimators. In this paper, we develop a computational approach to robust design for computer experiments without the need to assume independence or identical distribution of errors. Through explicit inclusion of the variance or correlation structures into the meta-model distribution, either maximum likelihood estimation or generalized estimating equations can be employed to obtain an appropriate Fisher information matrix. Robust designs can then be computationally sought which maximize some relevant summary measure of this matrix, averaged across a prior distribution of any unknown parameters.
\end{abstract}

\section{Introduction}
\label{sec:intro}
The fitting of a model to a sample of input-output data of a computationally expensive simulation is an important task in simulation analytics \citep{Santner_et_al_2003}. This model of a model (meta-model) can then be efficiently employed for prediction, optimization, or sensitivity analysis purposes. Sensitivity analyses are often well served by low-order polynomial (usually quadratic) meta-models, as they enable the characterisation of impact (main effects), synergies (two factor interactions) and diminishing returns (squared terms) \citep{Gill_et_al_2018,Sanchez_et_al_2012}. 

Fitting of the meta-model is enabled by a designed experiment, and for computationally expensive (often stochastic) simulations, the efficiency of the design is of importance \citep{fedorov_1972}. For linear meta-models, factorial-based designs (often fractional and supplemented with central and axial points if fully quadratic) are typically prescribed \citep{montgomery2012}, as these are efficient under $D$-optimality if the typically assumed condition of independent and identically distributed (iid) errors holds. 

\citet{Kleijnen_2015} is perhaps the seminal text on experimental design for simulation, and discusses the implications of departures from iid conditions for the analysis of linear meta-models, which \citet{Gill_2019} illustrates. However, while the assumption of independence can actually be assured in simulation by employing unique pseudo random number (PRN) streams at each design point, this overlooks an important variance reduction (design efficiency) opportunity. \citet{schruben1979} were the first to devise a design efficient PRN assignment strategy for linear meta-models and (generally) factorial-based designs (\citet{Gill_2021} illustrates with a simple example). 

However, \citet{Kleijnen_2015} is relatively silent on the question of design when iid conditions do not hold for linear meta-models (\textit{``the literature pays little attention to the derivation of alternative designs for cases with heterogeneous output variances"} and \textit{``the literature pays no attention to the derivation of alternative designs for situations with common random numbers (CRN)}"). Furthermore, simulation outputs are often discrete and sometimes only binary, so the broader range of generalized linear (meta-)models (GLMs) are typically required (i.e., linear, Poisson, and logistic) \citep{dunn2018generalized}. \citet{Woods_et_al_2006} point out that the design efficiency for GLMs depends on the regression parameters yet to be estimated, so that robust designs are often sought by computational optimization. 

In this paper, we seek to bring to the attention of the simulation analytics community literature which address some of these design-related gaps. In particular, we illustrate in some detail design construction for linear meta-models in the presence of heteroscedasticity (drawing on \citet{atkinson_cook_1995}) and GLMs in the presence of correlation \citep{Woods_Ven_2011}, before concluding with a proof of concept for the idea of jointly optimizing both the design and PRN assignment for linear meta-models.

\section{Robust design construction for GLM}
\label{sec:GLM_Designs}
\subsection{GLM designs}
In the GLM framework, for each input of $q$ factors $\mathbf{x}_i = [x_{i,1},x_{i,2}, \ldots, x_{i,q}]\in\mathbb{R}^{q}\quad i=1,\ldots,n$ we have a simulation response $Y_i$ with a probability mass/density function $p(y_i)$ assumed to come from the exponential family of distributions
and where there is an appropriate link function $g(\cdot)$ such that $g\left(\Esub{Y_i|\mathbf{x}_i}{\bvec{Y}}\right) = \bvec{f}^\text{T}(\mathbf{x}_i)\boldsymbol{\beta}$. Here $\boldsymbol{\beta}=[\beta_0, \beta_1, \ldots, \beta_{d-1}]^\text{T}$ is a column vector of $d$ $(q<d\leq n)$ unknown parameters and $\bvec{f}^\text{T}(\mathbf{x}_i):\mathbb{R}^{q}\to\mathbb{R}^{d}$ is a row vector of $d$ terms that may include first order and higher order interactions of the $q$ input factors. 

The goal is to choose the set of $n$ design points $\bvec{X} = [\mathbf{x}_1,\mathbf{x}_2, \ldots, \mathbf{x}_n]^\text{T}\in\mathbb{R}^{n\times q}$ to efficiently estimate $\boldsymbol{\beta}$. Minimizing the approximate volume of the covariance ellipsoid of the maximum likelihood estimator of $\boldsymbol{\beta}$ is equivalent to maximizing the determinant of the Fisher information matrix (hence called $D$-optimal)
\begin{equation}
\bvec{X}^* = \argmax_{\bvec{X}} | I_{\bvec{X}}(\boldsymbol{\beta})|, \quad I_{\bvec{X}}(\boldsymbol{\beta})_{j,k} = -\CondEsub{\pddydxdz{\ell(\bvec{Y},\mathbf{F}, \boldsymbol{\beta})}{\beta_j}{\beta_k} }{\boldsymbol{\beta}}{\bvec{Y}}
\label{eq:Dopt}
\end{equation} 
where $\mathbf{F} = [\bvec{f}^\text{T}(\mathbf{x}_1),\bvec{f}^\text{T}(\mathbf{x}_2), \ldots, \bvec{f}^\text{T}(\mathbf{x}_n)]^\text{T}$ is an $n\times d$ matrix and $\ell(\bvec{y},\mathbf{F}, \boldsymbol{\beta})=\sum_{i=1}^n\log p(y_i|\bvec{f}^\text{T}(\mathbf{x}_i,\boldsymbol{\beta}))$ is the (assumed twice differentiable) log-likelihood for the observations $\bvec{y} = [y_1,y_2,\ldots,y_n]^\text{T}$ at the design points $\bvec{X}$ given the parameters $\boldsymbol{\beta}$. 

Using second order partial derivatives of $\log p(y_i|\bvec{f}^\text{T}(\mathbf{x}_i),\boldsymbol{\beta})$ with respect to $\boldsymbol{\beta}$, then taking expectations with respect to $Y_i$, we obtain the following expression for the expected Fisher information matrix
\begin{equation}
I_{\bvec{X}}(\boldsymbol{\beta}) =  \mathbf{F}^{\text{T}}\bvec{P}\mathbf{F} 
\label{eq:obsFIM}
\end{equation}
where $\bvec{P} = \diag {\left(1/\left[g'(\Esub{Y_i|\mathbf{x}_i}{\bvec{Y}})^2\CondVarsub{Y_i}{\bvec{x}_i}{\bvec{Y}}\right]\right)}$ which  is a known function of $\mathbf{F}$ and $\boldsymbol{\beta}$ for the relevant exponential family distribution using in the GLM. 

Designs based on (\ref{eq:Dopt}) and (\ref{eq:obsFIM}) assume a fixed number of design points $n$ (an exact design). If instead we ascribe to $\bvec{x}_i$ a weight $0\leq w_i\leq 1$ (with $\sum_{i=1} w_i = 1$ thus representing how sampling effort is distributed across design points) and relabel $\bvec{x}_i = [x_{i,1},x_{i,2},\ldots,x_{i,k},w_i]$, so that $\bvec{X}\in\mathbb{R}^{n\times q+1}$, 
then the approximate design problem is to find $\bvec{X}^* = \argmax_{\bvec{X}} | I_{\bvec{X}}(\boldsymbol{\beta})|$ where $I_{\bvec{X}}(\boldsymbol{\beta}) = \mathbf{F}^{\text{T}}\bvec{W}\bvec{P}\mathbf{F}$
where $W_{ii}=w_i$. From this approximate design, an exact design of a particular size can be generated by sampling according to the weights $w_i^*$. 

\subsection{Robust design}
\label{sec:Robustdesign}
Obviously the requirement to know $\boldsymbol{\beta}$ \textit{a priori} is not useful for finding designs for estimating $\boldsymbol{\beta}$. A common approach to remove the $\boldsymbol{\beta}$ dependency is to average (some monotonic function of) the optimality criterion across a prior distribution $\pi(\boldsymbol{\beta})$ of possible values of $\boldsymbol{\beta}$
\begin{equation}
\bvec{X}^*=\argmax_{\bvec{X}} \int \log\left(|I_{\bvec{X}}(\boldsymbol{\beta})|\right) \pi(\boldsymbol{\beta})\, \text{d} \boldsymbol{\beta}
\label{eq:robustDopt}
\end{equation}
where the logarithm of the determinant is often used for numerical stability purposes. We call this pseudo-Bayesian approach \citep{Chaloner1995,Englezou2018} robust design, as it is robust to misspecification of the parameters (though not the meta-model - see Section~\ref{sec:future_research}). The prior can be based on previous investigations or subject matter expertise, or a non-informative probability distribution if required. 

Often, the integral in (\ref{eq:robustDopt}) is not analytically tractable, so numerical integration is required. Quadrature rules are possible but are more cumbersome in higher dimensions, so here we use a direct Monte Carlo estimator, so that
\[
\bvec{X}^*\approx\argmax_{\bvec{X}}  \frac{1}{M}\sum_{m=1}^M \log\left(| I_{\bvec{X}}(\boldsymbol{\beta}_m)|\right),
\label{eq:mc}
\]
where $\boldsymbol{\beta}_1, \boldsymbol{\beta}_2, \ldots, \boldsymbol{\beta}_M$ are iid draws from the prior $\pi(\boldsymbol{\beta})$.  

Robust design using a Monte Carlo estimate of the expected Fisher information requires the maximization of a random quantity with variance of $\mathcal{O}(1/M)$. Many standard non-linear optimization algorithms, such as Levenberg-Marquardt~\citep{Levenberg1944,Marquardt1963}, cannot handle random variables in the function to be optimized. Instead, we apply simulated annealing, which is a probabilistic optimization technique~\citep{Kirpatrick1983}. Other methods for stochastic optimization such as the Approximate Coordinate Exchange algorithm~\citep{Overstall2017} can be more efficient for more complex functions, but simulated annealing is sufficient here.

\section{Robust designs for departures from iid conditions}
\label{sec:GLM_Designs_IID}
\subsection{Linear meta-model in the presence of heteroscedasticity}
Consider a $q=2$ design problem $\bvec{x} = [x_1,x_2]\in [-1,1]^2$ for the full second-order polynomial linear meta-model, so $\bvec{f}^\text{T}(\mathbf{x}_i)=[1,x_{i,1},x_{i,2},x_{i,1}x_{i,2},x_{i,1}^2,x_{i,2}^2]$ and $g(\cdot)$ is the identity function, but where $Y_i\sim N(\bvec{f}^\text{T}(\mathbf{x}_i)\boldsymbol{\beta},\sigma^2v(\bvec{x}_i))$ with $v(\bvec{x}_i) = \exp(\bvec{x}_i\boldsymbol{\alpha}[1+2\bvec{x}_i\boldsymbol{\alpha}])$. Here  $\boldsymbol{\alpha}$ is a column vector with the same dimension as $\bvec{x}_i$ but otherwise unknown. Thus, the $Y_i$ are independent, but  $||\boldsymbol{\alpha}||^2_2$ controls the degree of heteroscedasticity. 

To find robust designs, we need the expected information matrix for this meta-model. Since $Y_i$ is normally distributed, it's log-likelihood at design point $\bvec{x}_i$ is
\[
\ell_i(y_i,\bvec{f}^\text{T}(\mathbf{x}_i),\boldsymbol{\beta},\boldsymbol{\alpha})=\frac{-(y_i-\bvec{f}^\text{T}(\mathbf{x}_i)\boldsymbol{\beta})^2}{2\sigma^2\exp(\bvec{x}_i\boldsymbol{\alpha}[1+2\bvec{x}_i\boldsymbol{\alpha}])}-\frac{\bvec{x}_i\boldsymbol{\alpha}[1+2\bvec{x}_i\boldsymbol{\alpha}]}{2}-\log(\sqrt{2\pi}\sigma)
\]
and it is relatively easy to show that $\mathbb{E}_{Y_i}\left[\frac{\partial^2\ell_i}{\partial\beta_j\partial\alpha_k}\right]=0$ given $\Esub{Y_i|\mathbf{x}_i}{\bvec{Y}}=\bvec{f}^\text{T}(\mathbf{x}_i)\boldsymbol{\beta}$, which means the Fisher information matrix in (\ref{eq:Dopt}) will be block diagonal with two blocks; one for $\boldsymbol{\beta}$ and the other for $\boldsymbol{\alpha}$. For these
\begin{eqnarray*}
\frac{\partial^2\ell_i}{\partial\beta_j\partial\beta_k} & = & \frac{-\bvec{X}_{i,j}\bvec{X}_{i,k}}{\sigma^2\exp(\bvec{x}_i\boldsymbol{\alpha}[1+2\bvec{x}_i\boldsymbol{\alpha}])} \\
\frac{\partial^2\ell_i}{\partial\alpha_j\partial\alpha_k} & = & -\frac{1}{2}\bvec{X}_{i,j}\bvec{X}_{i,k}\left(4-\left[4-(1+4\bvec{x}_i\boldsymbol{\alpha})^2\right]\frac{(y_i-\bvec{f}^\text{T}(\mathbf{x}_i)\boldsymbol{\beta})^2}{\sigma^2\exp(\bvec{x}_i\boldsymbol{\alpha}[1+2\bvec{x}_i\boldsymbol{\alpha}])}\right).
\end{eqnarray*}

Clearly, $I_{\bvec{X}}(\boldsymbol{\beta})$ takes the form (\ref{eq:obsFIM}) with $\bvec{P}_{ii}=(\sigma^2\exp(\bvec{x}_i\boldsymbol{\alpha}[1+2\bvec{x}_i\boldsymbol{\alpha}]))^{-1}=(\sigma^2v(\bvec{x}_i))^{-1}$ and given $\Esub{(y_i-\bvec{f}^\text{T}(\mathbf{x}_i)\boldsymbol{\beta})^2|\mathbf{x}_i}{\bvec{Y}}=\sigma^2\exp(\bvec{x}_i\boldsymbol{\alpha}[1+2\bvec{x}_i\boldsymbol{\alpha}])$ we see that $I_{\bvec{X}}(\boldsymbol{\alpha})=\bvec{X}^{\text{T}}\bvec{Q}\bvec{X}$ with $\bvec{Q}_{ii}=\frac{1}{2}[1+4\bvec{x}_i\boldsymbol{\alpha}]^2$. We note that for linear meta-models, $\bvec{P}$ and $\bvec{Q}$ do not depend on $\boldsymbol{\beta}$. This accords with \citet{atkinson_cook_1995} and their original derivation which showed that the information expected to be obtained about $\boldsymbol{\beta}$ based on the $i-$th design point is given by $\bvec{f}^\text{T}(\mathbf{x}_i)\bvec{f}(\mathbf{x}_i)/(\sigma^2v(\bvec{x}_i))$ while for $\boldsymbol{\alpha}$ it is presented as $\bvec{J}^T\bvec{J}$, where $\bvec{J}=(1+4\bvec{x}_i\boldsymbol{\alpha})\bvec{x}_i/\sqrt{2}$.

\begin{figure}[htb]
    \centering
    \includegraphics[width=\linewidth]{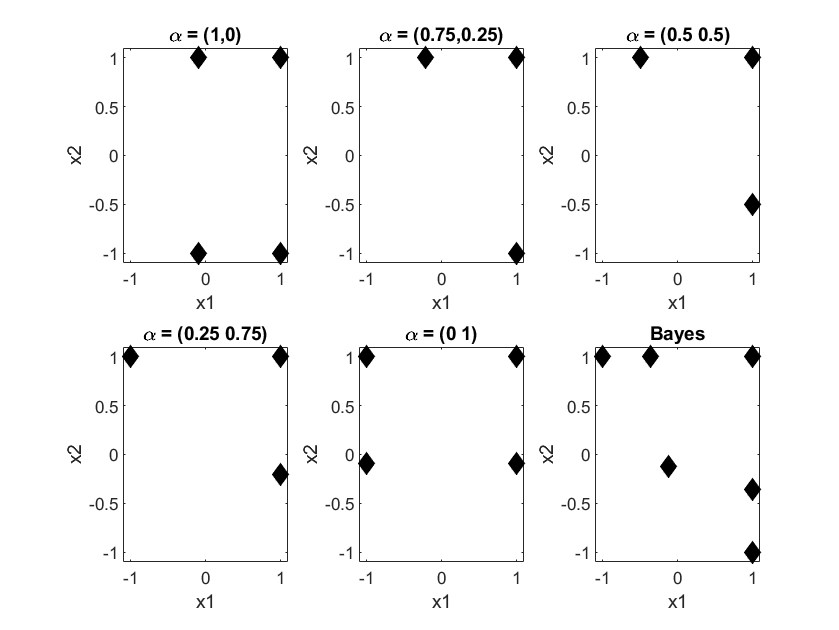}
    \caption{$D$-optimal designs for various values of $\boldsymbol{\alpha}$ and the robust (Bayes) design.}
    \label{fig:figure5}
\end{figure}

As a means of comparison,the prior considered in \citet{atkinson_cook_1995} for $\boldsymbol{\alpha}$ placed equal mass on the following fives values: $[1,0]$, $[0.75,0.25]$, $[0.5, 0.5]$, $[0.25, 0.75]$ and $[0,1]$.  The motivation is that these values span the directions in which the variance increases with $x_1$ and $x_2$, and that there is no prior knowledge to suggest which direction is more likely than another. 
Figure~\ref{fig:figure5} shows the local $D$-optimal designs for each unique value of $\boldsymbol{\alpha}$ along with the robust design, assuming the mean is known (thus focusing on $I_{\bvec{X}}(\boldsymbol{\alpha})$). 

Simulated annealing was employed to locate each design including the design weights. 
To do so, the optimisation was initialised with a random selection of design points and design weights with a relatively large value of $n$.
Throughout the optimisation, if some weights approached zero, then the corresponding design points were removed, which is why some optimal designs have different numbers of unique experimental runs.

Notably, these designs are very similar to those presented in Figure 5 of \citet{atkinson_cook_1995} (including the weights $w_i$, not shown here).  For the locally optimal designs, symmetry about $\boldsymbol{\alpha}$ is observed.  This is expected given how $\boldsymbol{\alpha}$ and $\bvec{x}$ exist in the model.  The robust design resembles a compromise between the designs found for each value of $\boldsymbol{\alpha}$ with the largest experimental effort being assigned to $\bvec{x}=[1,1]$.  Further, the points for $x_1=1$ and $x_2=1$ align with design points selected for different values of $\boldsymbol{\alpha}$.  Lastly, there is an inner point placed near $\bvec{x}=[0,0]$ which appears to be a compromise between the additional design point found at extreme values for $\boldsymbol{\alpha}$.

\subsection{Logistic meta-model in the presence of correlation}
\subsubsection{Fisher information matrix via generalized estimating equations}
Now consider a Bernoulli response $Y_i$, so that $P(Y_i=1)=p_i=\Esub{Y_i|\mathbf{x}_i}{\bvec{Y}}$ and $g(\cdot)$ is the logit function, with $q=3$ input factors and their pairwise interactions 
\begin{equation}
\text{logit}(p_i)=\beta_0+\beta_1x_{i,1}+\beta_2x_{i,2}+\beta_3x_{i,3}+\beta_4x_{i,1}x_{i,2}+\beta_5x_{i,1}x_{i,3}+\beta_6x_{i,2}x_{i,3}+\epsilon_i
\label{eq:logitGLMcor}
\end{equation}
but where we have added latent random variables $\boldsymbol{\epsilon}\sim N(\bvec{0},\bvec{R})$ with a general $n\times n$ covariance matrix $\bvec{R}_{i,j}=R(\bvec{x}_i,\bvec{x}_j)$. Unlike the linear meta-model, the $\text{logit}(\cdot)$ introduces non-linear terms into the expression for the log-likelihood, which will render the analytical integration over $\boldsymbol{\epsilon}$ to obtain the marginal likelihood impossible. Therefore, it is not possible to obtain an exact analytic expression to the expected Fisher information matrix for the model given in \eqref{eq:logitGLMcor}. 

However, following \citet{Woods_Ven_2011}, we can obtain an approximation using generalized estimating equations (GEE) (see \citet{Liang_Zeger_1986} for details). For the logistic GLM with correlations, the GEE leads to the following approximation (in the weighted design context)
\begin{equation}
I_{\bvec{X},\bvec{R}}(\boldsymbol{\beta}) \approx \mathbf{F}^{\text{T}}(\bvec{W}\bvec{P})^{1/2}\bvec{R}^{-1}(\bvec{W}\bvec{P})^{1/2}\mathbf{F},
\label{eq:obsFIMcor}
\end{equation}
where the dependence on $\boldsymbol{\beta}$ is observed through $\bvec{P}$ with $\bvec{P}_{ii}=p_i(1-p_i)=\exp(\bvec{f}^\text{T}(\mathbf{x}_i)\boldsymbol{\beta})(1+\exp(\bvec{f}^\text{T}(\mathbf{x}_i)\boldsymbol{\beta}))^{-2}$, and  $\bvec{W}_{ii}= w_i$ with weight $0\leq w_i\leq 1$ (with $\sum_{i=1} w_i = 1$ thus representing how sampling effort is distributed across design points) and relabel $\bvec{x}_i = [x_{i,1},x_{i,2},\ldots,x_{i,k},w_i]$, so that $\bvec{X}\in\mathbb{R}^{n\times q+1}$. 

For the purposes of this study, we assume constant (homoscedastic) variance $R(\bvec{x}_i,\bvec{x}_i)=\sigma^2$. The covariance structures we consider are as follows (for $i\neq j$).

\begin{itemize}
	\item \textit{Independent}: The standard assumption in which \eqref{eq:obsFIMcor} reduces to \eqref{eq:obsFIM}, i.e. $R(\bvec{x}_i,\bvec{x}_j)=0.$
	\item \textit{Constant}: All observations are equally correlated with each other, i.e. $R(\bvec{x}_i,\bvec{x}_j)=\sigma^2\rho.$
	\item \textit{Auto-regressive}: The observation index is treated as a time index, i.e. $R(\bvec{x}_i,\bvec{x}_j)=\sigma^2\rho^{|i-j|}.$
	\item \textit{Distance-kernel}: Isotropic spatial correlation between design points, i.e. $R(\bvec{x}_i,\bvec{x}_j)=\sigma^2\rho e^{-\frac{1}{4}\|\bvec{x}_i-\bvec{x}_j\|_2^2}.$
\end{itemize}

Here $\rho \in [0,1]$ is a correlation parameter (for now considering only positive correlations). Each correlation structure could, in principle, be valid for a specific computer simulation experiment. If this structure is know \emph{a priori}, then that structure should be used for the design. However, for most experiments, the correlation structure is not known. Therefore, we seek to understand the efficiency of each correlation assumption under misspecification.   

\subsubsection{Evaluating design efficiency under misspecification}
To consider the question of design efficiency for the logistic GLM with correlations \eqref{eq:logitGLMcor}, we perform a simulation study. For each of the above correlation structures we obtain a robust design using our computational approach, and assess the efficiency under misspecification of that correlation. We define some notation to express this comparison more formally. Let $\bvec{X}^*(\bvec{R})$ denote a robust design under the $D$-optimality criterion \eqref{eq:robustDopt} using the GEE approximation \eqref{eq:obsFIMcor} with covariance matrix $\bvec{R}$. Then define
\[
\mathscr{J}(\bvec{X},\bvec{R}) = \int\left|I_{\bvec{X},\bvec{R}}(\boldsymbol{\beta})\right|^{1/d}\pi(\boldsymbol{\beta})\text{d} \boldsymbol{\beta} \approx \frac{1}{M} \sum_{m=1}^M \left|I_{\bvec{X},\bvec{R}}(\boldsymbol{\beta}_m)\right|^{1/d}
\]
where the $d-$th root is routinely used to allow fair comparisons between designs. The ratio $\mathscr{J}[\bvec{X}_1,\bvec{R}]/\mathscr{J}[\bvec{X}_2,\bvec{R}]$ gives the $D$-efficiency of a design $\bvec{X}_1$ relative to a reference design $\bvec{X}_2$ given a covariance matrix $\bvec{R}$ ~\citep{Woods_Ven_2011}. The $D$-efficiency can be interpreted as the amount of additional experimental effort needed, whereby if $D$-efficiency is $0.5$, then you would need to run the design twice to obtain as much information as the optimal design. 

Now consider two covariance matrices $\bvec{R}_1$ and $\bvec{R}_2$, then $\bvec{X}^*(\bvec{R}_1)$ denotes the robust design assuming $\bvec{R}_1$, and similarly $\bvec{X}^*(\bvec{R}_2)$ is robust assuming $\bvec{R}_2$. It follows, that
\begin{equation}
\text{Misspecification } D\text{-Efficiency}(\bvec{R}_1,\bvec{R}_2) = \frac{\mathscr{J}[\bvec{X}^*(\bvec{R}_1),\bvec{R}_2]}{\mathscr{J}[\bvec{X}^*(\bvec{R}_2),\bvec{R}_2]},
\label{eq:effloss}
\end{equation}
represents the $D$-efficiency of a design using the misspecified $\bvec{R}_1$ when $\bvec{R}_2$ was the true covariance. 

\begin{figure}[htb]
	\centering
	\includegraphics[width=\linewidth]{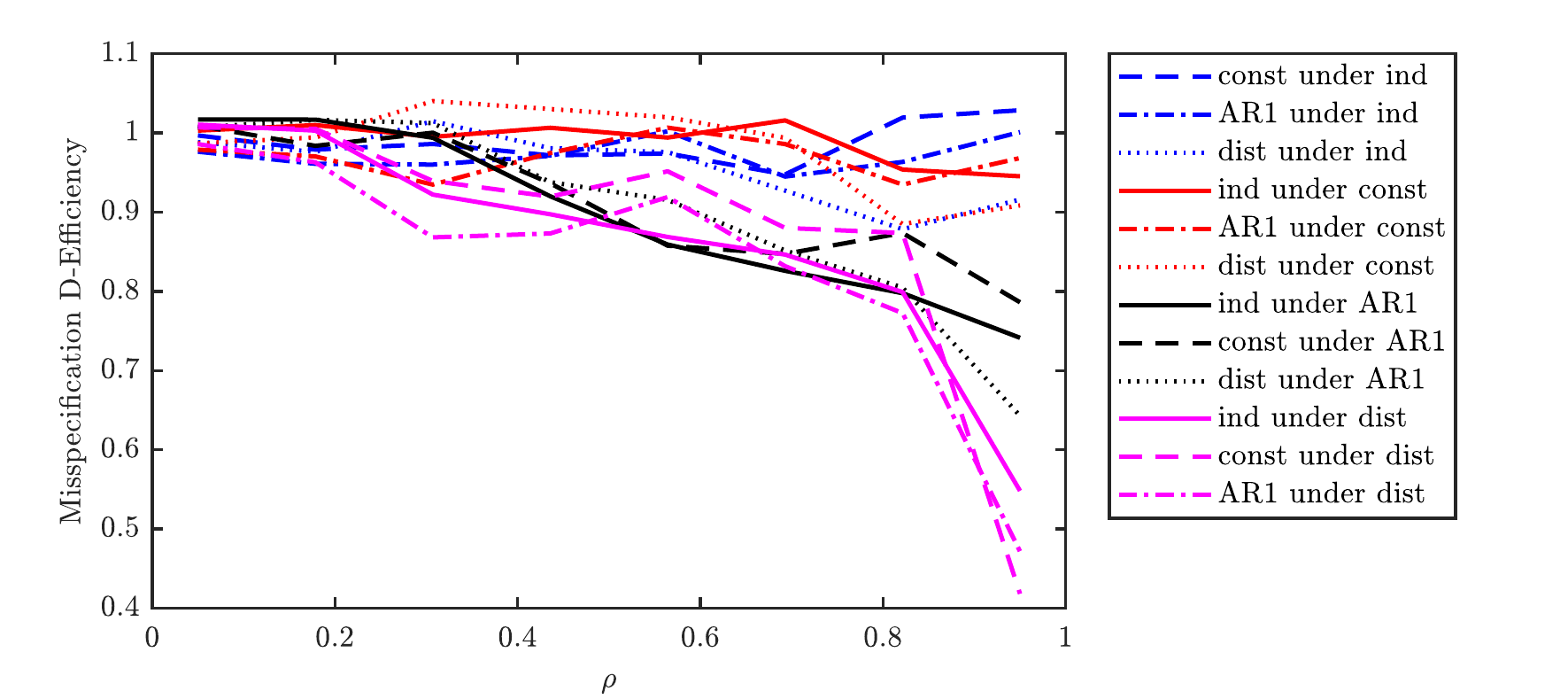}
	\caption{The efficiency of designs under different combinations of assumed ($\bvec{R}_1$) under true ($\bvec{R}_2$) correlation assumptions plotted against correlation strength.}
	\label{fig:fig1efficientb200init3n20}
\end{figure}

We evaluate this misspecification efficiency \eqref{eq:effloss} for each pair of covariance functions and do this for a range of $\rho \in [0.05,0.95]$ to investigate how the efficiency depends on the correlation strength. For each design simulation, we optimize $n$ weighted design points for the $q = 3$ factor model using the robust design expected utility estimated with $M = 1,000$ prior samples. When evaluating the final efficiency losses we use a more precise Monte Carlo estimate with $M= 20,000$. The resulting efficiency as a function of correlation strength is provided for each pair of covariance structures in Figure~\ref{fig:fig1efficientb200init3n20}. 

Note that the efficiencies $>1$ suggest too small a value of the Monte Carlo sampling rate ($M=1,000$). However, to produce Figure~\ref{fig:fig1efficientb200init3n20} required 25 robust designs (8 values of $\rho$ for each of the 3 covariance options dependent on $\rho$ with an additional independent case), with each design run costing approximately $16$ hours of CPU time on a Intel Xeon Gold 6140 processor (total of approximately $400$ CPU hours distributed over 18 cores) which motivated the chosen value of $M$ for this study.

Several important patterns are observed in Figure~\ref{fig:fig1efficientb200init3n20}. Firstly, there is almost no penalty for assuming a correlation structure when independence is valid. That is, the efficiency of constant, auto-regressive or distance correlation relative to independence is $> 0.9$ (Figure~\ref{fig:fig1efficientb200init3n20}, blue lines). A similar insensitivity is apparent for misspecification relative to constant correlation. However, the situation becomes quite different when considering misspecification relative to auto-regressive or distance-based correlation. In both cases, the penalty of misspecification increases as the correlation strength, $\rho$, increases. Assuming distance correlation when auto-regressive is true performs better overall that the converse relationship. However, when $\rho > 0.7$, constant correlation starts to be the better assumption under misspecification by auto-regressive or distance correlation.       

Thus, if sufficient knowledge is available to prescribe a correlation assumption with certainty, then this will always be the best choice. Beyond this unrealistic case, one clear result is that the independent assumption should only be used if the risks of misspecification is low. The same can mostly be said for the constant assumption, unless the correlation strength is higher, in which case the more specific structural distinctions between auto-regressive and distance correlation become apparent. The choice of auto-regressive or distance based correlation  assumptions do not reduce the quality of the design substantially if independent or constant correlations would have been also been valid choices. However, the choice of auto-regressive and distance correlation is more complex and depends on the correlation strength. If one can rule out auto-regressive correlation, that represents temporal correlation, then this causes few problems and distance correlation should be used. However, if it is unclear if distance or auto-regressive are possibilities, then additional exploration is needed. In general we arrive at the following recommendations.

\begin{enumerate}
	\item If $R(\bvec{x}_i,\bvec{x}_j)$ is known, use this in the design process.
	\item If $R(\bvec{x}_i,\bvec{x}_j)$ is uncertain, but auto-regressive correlation can be excluded, then use distance-kernel correlation.
	\item If $R(\bvec{x}_i,\bvec{x}_j)$ is completely uncertain, some understanding of the range of $\rho$ is required. If $\rho \leq 0.7$ then distance-kernel correlation is more robust, otherwise constant correlation is more robust.
\end{enumerate}       

\begin{figure}[htb]
		\centering
		\includegraphics[width=\linewidth]{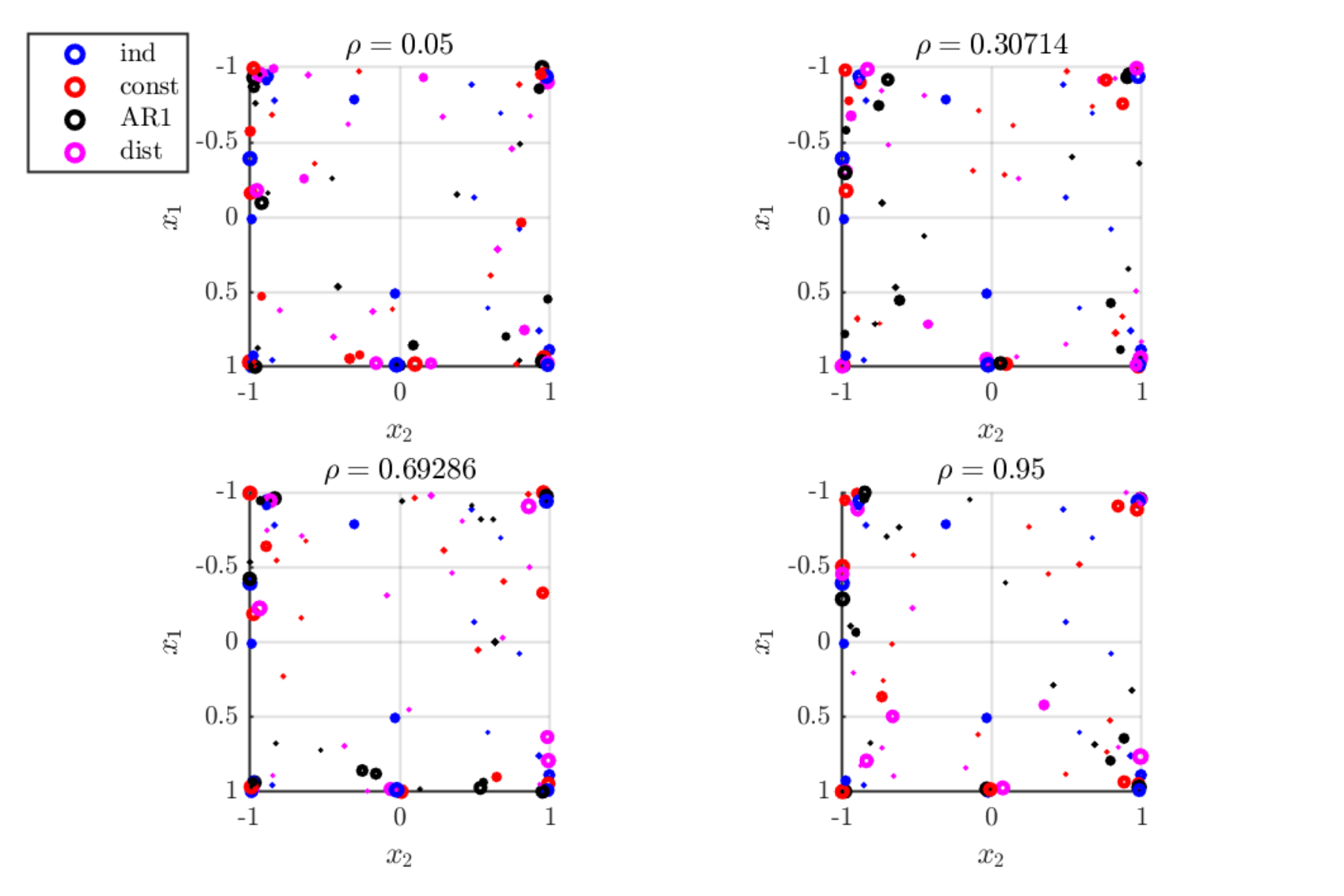}
		\caption{The weighted design points under different correlation assumptions. The weight of a design point is represented by the circle radius. View is along the $x_1$-axis.}
		\label{fig:fig2points2}
	\end{figure}

We also investigated the qualitative differences in the design patterns for the various correlation structures and values of correlation strength $\rho$. The example spatial patterns shown in Figure~\ref{fig:fig2points2} correspond to a view along the $x_1$-axes (the other axes views are very similar qualitatively). 

Given the efficiency results, it is not surprising that the design patterns look similar for different values of $\rho$. However, for a fixed $\rho$ we can observe some differences between designs under different correlation assumptions. Both the independent and constant correlation cases are characterised with fewer points, but with relatively constant weights, however, auto-regressive and distance correlation tend to have more points with lower weights.  

\section{Joint optimization of design and PRN assignment}
\label{sec:JointOpt}
Sections~\ref{sec:GLM_Designs} and \ref{sec:GLM_Designs_IID} describe design construction for GLMs where iid conditions may not be present. Unlike physical experiments, where dependence or correlation may arise due to unavoidable constraints and nuisance blocking effects need to be accounted for, simulation experiments can guarantee independence by using different PRN streams for each design point. However, simulation experiments can conversely induce correlations by the use of CRN. \citet{schruben1979} were among the first to clearly illustrate how doing so can improve the D-efficiency of a given design, and provided an assignment strategy for mostly factorial-based designs. Of interest in this section is the merging of that idea with the design construction approaches of the previous sections. 

\subsection{Linear meta-model in the presence of correlation}
Suppose there are $2g$ PRN streams given by $g$ streams (denoted $R_1,\dots,R_{g}$) and their antitheses (denoted $\bar{R}_1,\dots,\bar{R}_{g}$). Denote the $2g$ streams $R_1,\dots,R_{2g}$ with $R_{g+j} = \bar{R}_j$, for $j=1,\dots,g$ and let $b(R_j)$ be the random block effect associated with PRN stream $R_j$. Suppose now that design point $\bvec{x}_i$ is assigned PRN stream $R_{k(i)}$ for some assignment strategy $k(\cdot)$. Let $\bvec{Z}$ be the $n \times 2g$ matrix with 
$$\bvec{Z}_{ih} = \left\{ \begin{array}{ll}
1 & \mbox{if $R_h$ is used for experimental run $i$;}\\
0 & \mbox{otherwise,}
\end{array} \right.$$
and $\bvec{b} = \left[ b(R_1), \dots, b(R_{2g})\right]^T$ be the $2g \times 1$ column vector of unique random block effects. Then, $\boldsymbol{\gamma}=\bvec{Z}\bvec{b}=\left[b(R_{k(1)}), \dots, b(R_{k(n)})\right]^T$ is the column vector of $n$ random block effects as assigned in the experiment. Now $\mathrm{E}\left(\boldsymbol{\gamma}\right) = \bzero_n$ and $\mathrm{Var}\left(\boldsymbol{\gamma}\right) = \bvec{R}$, where $\bvec{R}$ is the $n \times n$ matrix with
$$R_{i,j} = \mathrm{Cov}\left(b(R_{k(i)}),b(R_{k(j)})\right) = \left\{
\begin{array}{ll}
\sigma^2 \rho_+ & \mbox{if $k(i) = k(j)$;}\\
- \sigma^2 \rho_- & \mbox{if $|k(i) - k(j)| = g$;}\\
0 & \mbox{otherwise,}
\end{array} \right.$$
where $\rho_->0$ and $\rho_+>0$ are unknown. 

Here $\rho_+$ and $-\rho_-$ are positive and negative correlations induced by using the same PRN stream or its antithesis for experimental points $i$ and $j$. Following \citet{schruben1979}, the model is
\begin{equation}
Y_i=\bvec{f}^\text{T}(\mathbf{x}_i)\boldsymbol{\beta}+\gamma_i+\epsilon_i
\label{eqn:model}
\end{equation}
where $\mathrm{E}\left(\bepsilon\right) = \bzero$ and $\mathrm{Var}\left(\bepsilon\right) = \sigma^2 \left(1 - \rho_+\right)\bvec{I}$. For a linear meta-model, we can use the Ordinary Least Squares estimator $\hat{\bbeta} = \left(\mathbf{F}^\text{T}\mathbf{F}\right)^{-1} \mathbf{F}^\text{T}\bvec{y},$ which can be shown to have variance (under (\ref{eqn:model}))
\[
\mathrm{Var}\left(\hat{\bbeta}\right) = \sigma^2 \left(\mathbf{F}^\text{T}\mathbf{F}\right)^{-1} \mathbf{F}^\text{T} \bvec{V} \mathbf{F} \left(\mathbf{F}^\text{T}\mathbf{F}\right)^{-1} \mbox{ where  } \bvec{V} = (1-\rho_+)\bvec{I} + \bvec{Z}\bvec{R}\bvec{Z}^T.
\]

\subsection{Optimal blocked designs}
Optimal design for blocked experiments has been considered previously (see, for example, Chapters 7 and 8 of \citet{GJ2011} or Chapter 15 of \citet{ADT2007}). What is different here is that there is positive correlation between elements of $\bb$ whereas these are usually assumed to be independent. Design specification here involves both the choice of $\bvec{X} = \left[\bvec{x}_1,\dots,\bvec{x}_n\right]^\text{T}$ and the PRN allocation $\bvec{k} = \left[k(1),\dots,k(n)\right]$ where $k(i) \in \left\{1,\dots, 2g \right\}$. 

While before we used the (determinant of the) Fisher information matrix, as the goal is to minimize the volume of the covariance ellipsoid, here we can directly and equivalently use the (log of, for numerical stability) determinant of $\mathrm{Var}\left(\hat{\bbeta}\right)$
\[
(\bvec{X}^*,\bvec{k}^*)=\argmin_{(\bvec{X},\bvec{k})} \left(\log |\mathbf{F}^\text{T}\bvec{V}\mathbf{F}| - 2 \log |\mathbf{F}^\text{T}\mathbf{F}| \right).
\]

However, $\bvec{V}$ depends on unknowns $\rho_-$ and $\rho_+$ through $\bvec{R}$. As before, we instead seek a robust design under a joint prior distribution $\pi(\rho_-,\rho_+)$ with domain $[0,1]^2$
\begin{equation}
(\bvec{X}^*,\bvec{k}^*)=\argmin_{(\bvec{X},\bvec{k})} \left(\int_0^1 \int_0^1 \log |\mathbf{F}^\text{T}\bvec{V}\mathbf{F}|\pi(\rho_+, \rho_-) \mathrm{d}\rho_- \mathrm{d}\rho_+ - 2 \log | \mathbf{F}^\text{T}\mathbf{F}|\right)  \label{eqn:crit}
\end{equation}
and as the integral in (\ref{eqn:crit}) is not typically available in closed form, it is evaluated here using a 2-dimensional Gauss-Legendre quadrature rule (see, for example, \citet{mvQuad})
$$(\bvec{X}^*,\bvec{k}^*)\approx\argmin_{(\bvec{X},\bvec{k})}\left( \sum_{m=1}^M \omega_{m} \log | \mathbf{F}^\text{T}\bvec{V}_{m}\mathbf{F}| - 2 \log | \mathbf{F}^\text{T}\mathbf{F}|\right) $$
where $\omega_{1},\dots,\omega_M$ are the quadrature weights and $\bvec{V}_{m}$ is $\bvec{V}$ evaluated at the corresponding quadrature nodes $(\rho^{(m)}_-,\rho^{(m)}_+)$. For illustrative purposes it suffices here to use a rudimentary joint optimization
\begin{equation}
(\bvec{X}^*,\bvec{k}^*)\approx\argmin_{\bvec{X}}\left(\min_{\bvec{k}}\left(\sum_{m=1}^M \omega_{m} \log | \mathbf{F}^\text{T}\bvec{V}_{m}\mathbf{F}|\right) - 2 \log | \mathbf{F}^\text{T}\mathbf{F}|\right) \label{eq:approx}
\end{equation}
where the inner minimization is performed by enumerating over all possible $\bvec{k}$ and the outer using a simple coordinate exchange algorithm \citep{MN1995}. 

\subsection{Proof of concept}
Suppose $n=10$, and there are $q=2$ inputs with $\bvec{X} = [-1,1]^2$, $\bvec{f}^\text{T}(\mathbf{x}_i) = \left[1, x_{i,1},x_{i,2},x_{i,1}x_{i,2},x_{i,1}^2, x_{i,2}^2\right]$ so that $d=6$, and there is $g=1$ PRN stream. The discrete set of values in the coordinate exchange algorithm is $\left\{-1, -0.9, -0.8, \dots, 0.8, 0.9,1\right\}$. A robust design (denoted $\left(\bvec{X}_R, \bvec{k}_R\right)$) is found via (\ref{eq:approx}) where the prior joint distribution for $\rho_-$ and $\rho_+$ used were independent uniform distributions. This is compared to the design (denoted $\bvec{X}_C$) found by minimizing (\ref{eq:approx}) but where the same PRN stream is used for all design points (i.e., CRN where $k(i)=1,\, i=1,\dots,n$), and to the design (denoted $\bvec{X}_I$) found by minimizing (\ref{eq:approx}) but with a different PRN stream for each design point (i.e., independent, which is equivalent to minimizing $- \log |\mathbf{F}^\text{T}\mathbf{F}|$ and is the standard $D$-optimal design). Note that these are the same comparisons as made by \citet{schruben1979}. 

Both the independent and CRN designs converged to the face-centered central composite design with center point, while the robust design had repeated points at $[-1,-1]$ and $[+1,+1]$ and did not utilise the center point. The minimum values of $\log|\mathrm{Var}\left(\hat{\bbeta}\right)|$ are $-9.1, -11.8$ and $-13.0$ for the independent, CRN and robust designs, respectively, thus demonstrating the benefit of inducing correlation over favouring independence and of the benefit of using a combination of common and antithetic random number streams. 

Finally, an interesting comparison is the performance of the independent and CRN designs under the optimal allocation of the two PRN streams ($R_1$ and $\bar{R}_1$). It turns out that $\log|\mathrm{Var}\left(\hat{\bbeta}\right)| = -11.8>-13.0$ in both cases. This demonstrate the utility of jointly optimizing over the design points $\bvec{X}$ and PRN assignment $\bvec{k}$, i.e. simply using a standard $D$-optimal design and then applying the optimal PRN assignment strategy to that design (as originally performed by \citet{schruben1979}) can be outperformed by joint optimization.

\section{Summary}
\citet{Kleijnen_2015} provides important guidance on how to analyse simulation experiments in the event of departures from the ubiquitous independent and identically distributed assumptions. Heteroscedasticity in simulation output is not uncommon, and it is potentially beneficial to induce dependence through the reuse of pseudo-random number streams to reduce the generalized variance of the meta-model parameter estimators. 

In this paper, we focus on the experimental design (vice analysis) aspects, and employed a computational approach to robust design for expensive computer experiments without the need to assume independence or identical distribution of errors in the meta-model to be developed. Through explicit modelling of the variance component for linear meta-models, the Fisher information was obtained within a maximum likelihood inference framework, while explicit modelling of the correlation structure for generalized linear meta-models and  generalized estimating equations can be employed to approximate the Fisher information matrix. In both cases, robust designs can then be computationally sought which maximize some relevant statistic of this matrix, averaged across a prior distribution of any unknown parameters. 

Moving away from the assumption of independence implies that a correlation structure be introduced, the misspecification of which could have a negative effect on the performance of the design. We built upon \citet{Woods_Ven_2011} to begin investigation of robust designs for GLMs with correlations. However, our work is distinct to \citet{Woods_Ven_2011} as we investigated the effect of covariance matrix misspecification for a variety of correlation structures, in the context of a $3$-factor logistic GLM with pairwise interactions. While our results are not exhaustive, the cases of constant correlation, auto-regressive correlation, and distance correlation represent major classes of correlation structures (uniform, temporal, spatial) and are helpful to inform some recommendations.

As illustrated in Section~\ref{sec:JointOpt}, it may be effective to consider $\rho$ as part of the vector of unknown parameters and hence integrate over a joint prior probability density. The choice to look at the efficiencies of designs as a function of $\rho$ was primary to identify any dependencies between the effect of misspecification and the correlation strength. Since we mainly observe this dependency for the  auto-regressive cases, robust design over $\rho$ may only be required if  auto-regressive is a feasible correlation structure. It is also important to note that this simulation approach is designed to obtain some heuristics for dealing  with correlation assumptions. In practice, the $D$-efficiency for a real problem will never be available. However, the results provide some means to assist in the interpretation of confidence regions that are obtained for a design. That is, one must assume some misspecification and therefore treat predicted parameter uncertainty estimates as underestimates for the true uncertainty that could arise when the computer experiment is performed. 

Finally, \citet{schruben1979} pioneered the search for effective assignment strategies of pseudo-random numbers to design points, but did so with fixed (textbook) designs (and for linear meta-models only). In this paper, we provide an example proof of concept of the possibility of jointly optimizing the design and pseudo-random number assignment and show that gains in statistical efficiency can be made. 

\section{Future research}
\label{sec:future_research}
This paper has assumed the vector of covariates, representing the simulator inputs and configuration settings, are continuous. However, discrete covariates must also be dealt with. Challenges arise in this case since the structure of covariances can be more complex. Furthermore, stochastic optimization is substantially more challenging to deal with in the discrete covariate case. While simulated annealing can deal with discrete spaces~\citep{Kirpatrick1983}, it will be more computationally intensive. Unfortunately, methods like  Approximate Coordinate Exchange \citep{Overstall2017} can only deal with continuous design spaces, however, other methods may exist to handle a discrete design space \citep{MN1995}. Further work is needed to determine the most effective computational scheme for this case.

Model misspecification is a broad challenge in robust design for computer experiments. While accounting for heteroscedasticity and correlations improves the situation substantially, there is still the potential for bias in the design due to the meta-model being unable to replicate some behaviours of the complex computer model. One approach to deal with this is the inclusion of an additional discrepancy term using Gaussian processes \citep{Englezou2018,Kennedy2000}. 

For the joint optimization of the design and PRN assignment, the coordinate exchange algorithm used relied on complete enumeration of all possible $\bvec{k}(\cdot)$ assignments. The size of this set grows fast with the number of PRN streams $g$ and would appear difficult to apply even for $g>1$ and is therefore not particularly scalable. A more sophisticated approach will be required.

For non-linear meta-models we have focused on the logistic GLM that corresponds to a binary outcome from a simulation. However, the computational approach we consider here for robust design would also be applicable to other GLMs of interest, such as binomial and Poisson responses. For the joint optimization problem, a generalized linear mixed model (GLMM) approach might be applicable (see Chapter 17 of \citet{P2013}). The meta-model would then have the form $g\left(\Esub{\bvec{Y}|\mathbf{X}}{\bvec{Y}}\right) = \mathbf{F}\boldsymbol{\beta}+\boldsymbol{\gamma}$, where $\boldsymbol{\gamma}=\bvec{Z}\bvec{b}$ as in Section~\ref{sec:JointOpt}. The unknown parameters $\boldsymbol{\beta}$ could be estimated via maximum likelihood where 
$\mathrm{Var}(\hat{\boldsymbol{\beta}}) \approx \left( \mathbf{F}^T \bvec{M} \mathbf{F} \right)^{-1}$ and there are various forms for $\bvec{M}$ under different approximations \citep{P2013} . This modelling approach is an example of a GLMM, for which optimal design have been considered previously (see \citet{XS2021} and references therein). 


\bibliographystyle{wsc}

\end{document}